\documentstyle[12pt]{article}
\def\be{\begin{equation}}
\def\ee{\end{equation}}
\def\barr{\begin{array}}
\def\earr{\end{array}}
\def\ie{ {\it i.e.} }

\def\etal{ {\it et al.} }
\def\gsim{\:\raisebox{-0.5ex}{$\stackrel{\textstyle>}{\sim}$}\:}
\def\rp{$R_p \hspace{-1em}/\;\:$}
\begin{document}
\setcounter{page}{0}
\renewcommand{\thefootnote}{\fnsymbol{footnote}}
\thispagestyle{empty}
\vspace*{-1in}
\begin{flushright}
CERN-TH.7509/94 \\[2ex]
{\large \bf hep-ph/9412259} \\
\end{flushright}
\vskip 45pt
\begin{center}
{\Large \bf \boldmath $R$--Parity Violating SUSY
or Leptoquarks: \\[2ex]
Virtual Effects in Dilepton Production} \\
\vspace{11mm}
\bf
Gautam Bhattacharyya\footnote{ gautam@cernvm.cern.ch},
Debajyoti Choudhury\footnote{debchou@surya11.cern.ch}
and {\bf K. Sridhar\footnote{sridhar@vxcern.cern.ch} }\\
\vspace{13pt}
{\it Theory Division, CERN, \\ CH-1211, Geneva 23, Switzerland.}

\vspace{50pt}
{\bf ABSTRACT}
\end{center}

\begin{quotation}

In the Standard Model (SM), dilepton production in hadron--hadron collisions
proceeds through the conventional Drell--Yan mechanism
$q \bar{q} \rightarrow l^+ l^-$ with the exchange of a gauge boson.
 Some extensions of the SM contain a quark--lepton contact
interaction {\it via} a $q l \phi$ Yukawa coupling, where
$\phi$ is a scalar.
Theories with scalar leptoquarks and $R$--parity violating SUSY models
are the most important examples of such extensions.
These Yukawa couplings induce a
different dynamical configuration compared to the SM
 ($t$-channel vs. $s$-channel) in the $q\bar{q} \rightarrow
l^+l^-$ process
and thus offer the possibility of being
identifiable upon imposition of suitable kinematic cuts.
We discuss these effects in the context of the dilepton production
in the CDF experiment,
and explore consequences in the forthcoming Large
Hadron Collider (LHC).
\end{quotation}

\vspace{70pt}
\noindent
\begin{flushleft}
CERN-TH.7509/94\\
November 1994\\
\end{flushleft}

\vfill
\newpage
\setcounter{footnote}{0}
\renewcommand{\thefootnote}{\arabic{footnote}}
\setcounter{page}{1}
\pagestyle{plain}
\advance \parskip by 10pt

In the quest of understanding physics at energy scales beyond that of the
standard model (SM),
many ideas have been discussed rather extensively in the
literature. Two of the most popular, and in some sense intertwined,
theories are those of  grand unification and
supersymmetry. In both theories, there
arise new particles that connect the quark and lepton sectors.
Such a particle can either be a scalar or a vector. In this letter, we
shall restrict ourselves
to scalars. As for  weak $SU(2)$ properties, it can
transform either as
a singlet, a doublet or a triplet.

In grand unified theories \cite{gut},
where quarks and leptons are part of the same multiplet, there
naturally arise scalar leptoquarks leading to a
contact interaction between a quark and a lepton.
In the Minimal Supersymmetric Standard Model (MSSM) \cite{mssm}
too, the squarks may
mediate interactions between quarks and leptons unless some extra
symmetry ($R$--parity) is imposed on the theory. As there is no deep
motivation
for such a symmetry to actually exist, it is interesting to examine
the possible consequences of such $R$--parity violating (\rp)
 interaction. Within the context of the MSSM,
the part of the  Lagrangian violating $R$--parity may be written in
terms of the chiral superfields as
\be
    {\cal L}_{\not R} = \lambda_{ijk} L_i L_j E^c_k +
                        \lambda'_{ijk} L_i Q_j D^c_k +
                        \lambda''_{ijk} U^c_i D^c_j D^c_k ,
      \label{R-parity}
\ee
where $L_i$ and $Q_i $ are the $SU(2)$ doublet lepton and quark fields
and $E^c_i, U^c_i, D^c_i$
are the singlet  superfields. Concentrating on the
$\lambda'_{ijk}$ piece above, it is obvious that the phenomenology due
to such terms is very similar to that of certain scalar
leptoquarks.

A generic scalar coupling between a lepton ($l$) and a quark ($q$)
 can  be parametrized as
\be
   {\cal L} = g \bar{l} (h_L P_L + h_R P_R) q \phi +
              g \bar{l^c} (h_L' P_L + h_R' P_R) q \phi'  +
   \mbox{\rm h.c.},
     \label{generic Lagrangian}
\ee
 where $g$ is the weak gauge coupling constant,
$l^c$ is the charge conjugate
field and $P_{L,R} = (1 \mp \gamma_5)/2$ ($\phi$ and $\phi'$ are
independent fields).
Since diquark couplings violate baryon number,
non-observation of  proton decay forbids the simultaneous presence of
diquark couplings as well.\footnote{This implies that in
the case of
\rp~theories $\lambda''$ cannot simultaneously exist with
either of $\lambda$ or $\lambda'$. However,
$\lambda$ and $\lambda''$ are
not relevant to the rest of our discussion.}
Furthermore, to suppress
flavour-changing neutral current processes \cite{leurer},
 each such scalar is assumed
to couple primarily to only one family of quarks (leptons).
Finally, to
survive stringent bounds coming from helicity suppressed $\pi
\rightarrow e \nu$ and similar processes \cite{shank}, the
scalar  coupling
is assumed to be chiral ({\it i.e.}
 it is not allowed to simultaneously couple
to the left- and right-handed quarks)\footnote{In principle, couplings
of scalars with the left-handed quark doublets
cannot be made fully flavour diagonal in the quark sector,
since the presence of CKM mixings prevents its coupling to be diagonal
in the up- and down-sectors simultaneously.}.
 There is a subtlety in the case of the MSSM as the squarks
$\tilde{f}_L$ and $\tilde{f}_R$ will always mix. One may argue that such
effects are proportional to $m_f/m_{SUSY}$
and hence would be small in most cases
including the one that we are about to discuss.
Still, the existence of this mixing does affect some
rare processes, and,
as a result, the
bounds \cite{barger} on \rp\  couplings are somewhat stronger than
those for a chirally coupling leptoquark. With the above
 assumptions then, there
are only five possible scalar couplings involving a charged lepton and these
are listed in Table \ref{table}. We have grouped scalars with different
$SU(2)_L \otimes U(1)_Y$ properties together as their gauge interactions
are irrelevant for the processes that we intend to study.

Leptoquark couplings as well as \rp\ couplings
 are constrained by direct searches at various
 colliders \cite{collider}.
The Fermilab CDF \cite {cdflq} and D0 \cite{d0lq} experiments rule
out
a `first generation scalar' upto $\sim 100$ GeV, almost
irrespective of the size of the Yukawa couplings.
On the other hand, the HERA
collaboration
sets a limit of $\sim 180$ GeV \cite{hera} on the mass of a scalar
coupling with electroweak strength to
a first generation quark and electron.
Similar bounds also hold for the \rp\ case. Due to the presence
of other decay channels for the squarks, the exact constraints
are somewhat different\cite{butter}.
For example, the Tevatron dilepton \cite{dp}
data have been used
to set lower limits on squark/gluino masses for almost any value of
the \rp\ coupling.
For scalars coupling to the third generation, the bounds are
understandably weaker. The strongest constraints to date can be
inferred from the loop effects
on the leptonic partial widths of the $Z$ \cite{bes}.
Stronger constraints on such scalars can be obtained in future colliders
by looking either at $\tau$--number violating processes \cite{probir} at
LEP200
or at $t\bar{t}$ production \cite{dc} at the Next Linear Collider.

It is obvious that such experiments (other than those at HERA) have
very little to say about the Yukawa couplings {\it per. se.}
In this letter,  we point out an experiment that would do so.
To do this, we investigate the
possibility of identifying the effects of a virtual scalar
exchange in dilepton production in
hadronic reactions. The signature we focus on  is a
lepton pair {\em without} any missing transverse momentum. The lowest
order SM process leading to such a final state is the
Drell--Yan mechanism \ie $q \bar{q} \rightarrow (\gamma^\ast,
Z^\ast) \rightarrow l^+l^-$. The presence of such a scalar would
introduce an additional $t$--channel diagram, an obvious consequence
being a modification of the total cross section \cite{rz}.
A more sensitive probe,
though, could be a comparison of the differential distributions.
Owing to the extra diagram being a $t$--channel one in contrast to
the $s$--channel SM contribution, it is conceivable that the effect of the
former would be more pronounced for certain phase space configurations.

A particular distribution {\em viz.} the rapidity dependence suggests itself.
In the subprocess centre-of-mass (c.m.)
frame, one expects the $t$--channel contribution to
be more central than the $s$--channel one. Another
differential variable of interest could be  the invariant mass of the
lepton pair. Indeed, the CDF collaboration has studied
the latter  distribution for both dielectron and dimuon
pairs \cite{cdfdy}. In the first part of this letter, we show how this
experiment can discriminate between the SM and a theory with an extra
scalar interaction. We
compare the theoretical predictions with the existing data and point out
the improvements necessary to make our conclusions more quantitative.
In the second part, we examine the
same effect for the LHC, and speculate on the sensitivity that can be
achieved.

In principle, the contribution of a scalar of the type $\phi'$
(see eq. (\ref{generic Lagrangian})) can be distinguished from that of
type $\phi$ only if ($i$) the colliding beams are asymmetric
and ($ii$) the
individual lepton charges are distinguished. We shall not
dwell upon this possibility
here. We make the simplifying assumption
that at best one of the scalars in Table~\ref{table} is present.
As an example, we present here the
differential cross section for the process
$q \bar{q} \rightarrow l^+ l^-$ for the case $h_R = 0, \
h_L \neq 0$. This corresponds to scalars of Types $III,\ IV$  or $V$
in Table~1. (The expressions for the other types of scalars can be
obtained by simple modifications of the couplings.)
Defining the
 left- and right-handed couplings of the $Z$ to a fermion ($f$) by
\begin{equation}
a_L^f = (t_3^f - e_f \sin^2\theta_W)/{\sin\theta_W \cos\theta_W}, \qquad
a_R^f = - e_f \cot \theta_W\ ,
\end{equation}
where $e_f$ and $t_3^f$ are the corresponding charge and isospin
respectively
and $\theta_W$ is the weak mixing angle,
we have, for a scalar of mass $m_\phi$:
\be
\frac{{\rm d} \sigma } { {\rm d} t} =
\frac{{\rm d} \sigma_{\rm SM} } { {\rm d} t} +
   \frac{\pi g^2  t^2 }{3 M^6} K_t
      \left[ e_q e_l + a_L^q a_R^l
                  \frac{ M^2 (M^2-m_Z^2)}{(M^2-m_Z^2)^2+\Gamma_Z^2 m_Z^2}
  +        \frac{1}{4} K_t M^2
    \right],
\label {dsig}
\ee
where
\be
   K_t \equiv \frac{|h_L|^2}{t - m_\phi^2}
\ee
and $M$ is the invariant mass of the $l^+l^-$ pair.
The QCD corrections to the Drell--Yan process
have been calculated \cite{QCD} to the next-to-leading order and is
a function of the c.m. energy ($\sqrt{s}$) of the collider, the
structure functions used and the subprocess scale $M$.
For $M \gsim 20 $ GeV,
the dependence on $M$ is marginal and one may  approximate
it by a scale--independent constant.
In the absence of an explicit calculation of the higher-order effects,
we assume that
the QCD correction for the $t$-channel
scalar exchange process
is the same as that for the
Drell--Yan process.
One may argue that in the
presence of scalars, these corrections are likely to be even
larger.

To obtain the cross section for a
hadronic collision
$A B \rightarrow l^+l^- X$, we have to convolute the expression in
eq.~(\ref{dsig})
with the parton distributions. For example, the invariant mass
distribution is given by
\be
{d\sigma \over {dM}} = {{2M}\over s}
       \sum_q \int dx dt \left[q_A(x) \bar{q}_B(M^2/s x)
{d\sigma\over dt}  + (A\leftrightarrow B)\right],
                 \label{convolution}
\ee
where $x$ is the fraction of the momentum of $A$ carried by $q_A$.
While the sum in eq.(\ref{convolution}) runs
over all species of quarks for the SM piece,
for the scalar contribution
it runs only over those quarks
with which the scalar couples. For the rest
of this analysis, we shall
concentrate only on those that couple to one or
both of the first generation quarks.
We use the MRSD-$'$ structure functions \cite{mrs}
evaluated at the scale $Q^2 = M^2$.
For this choice, the QCD $K$-factor $\simeq 1.3 (1.1) $ for
$\sqrt{s} = 1.8 (14)$ TeV.

\begin{figure}[htb]
\vskip 8in\relax\noindent\hskip -1in\relax{\includegraphics{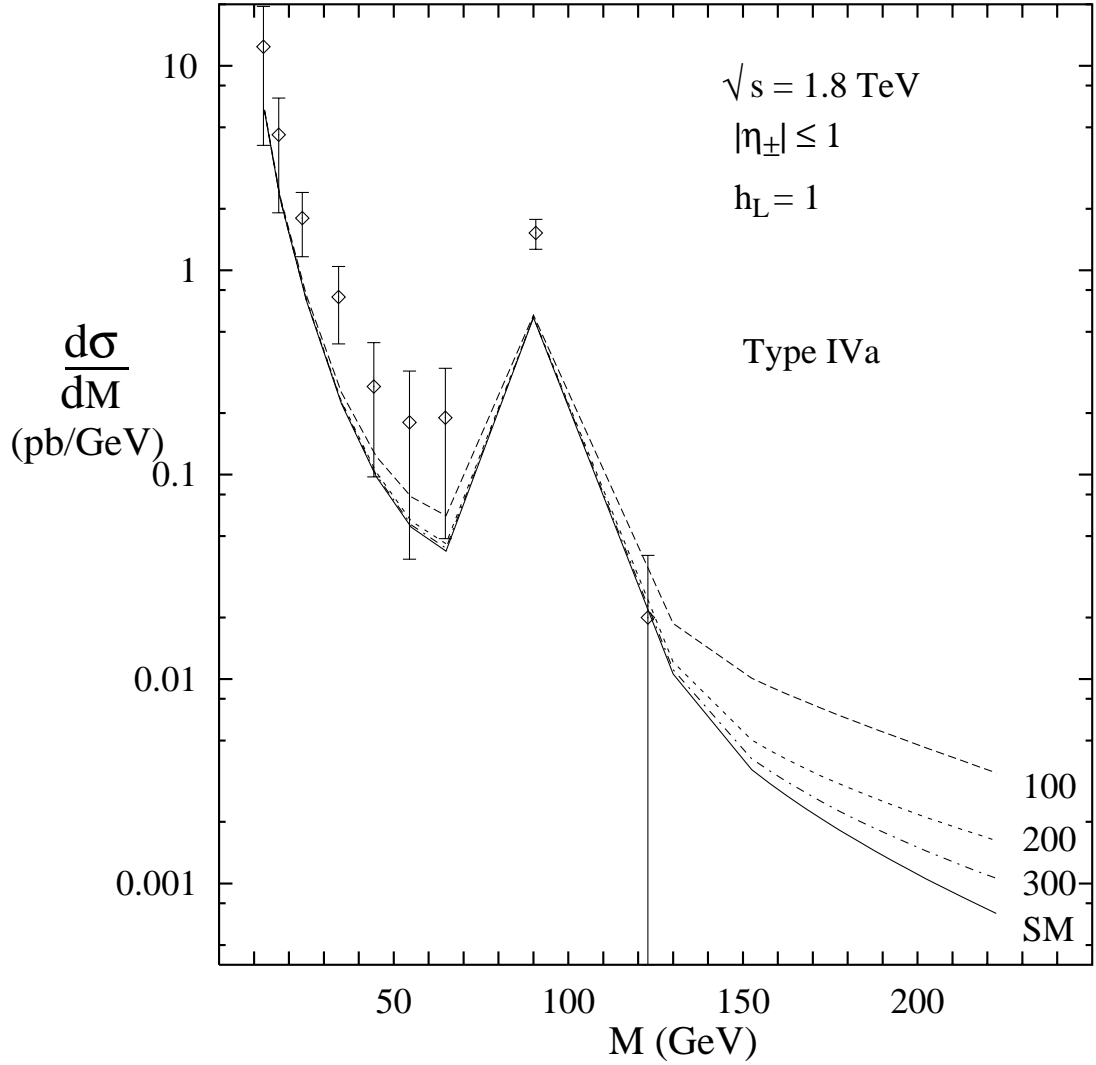}}

\vspace{-20ex}
\caption{The dilepton invariant mass distribution at CDF. The curves
correspond to the theoretical expectation for the SM and for leptoquarks
(Type $IVa$) of masses $m_\phi = 100,\:200,\:300$ GeV respectively.
For $M \leq 150$ GeV,
the invariant mass binning is the same as that used in
\protect\cite{cdfdy}.
}
       \label{cdffig}
\end{figure}

The CDF collaboration at Fermilab measures the Drell--Yan
cross section \cite{cdfdy}
$d\sigma/dM$ in the range $11 < M < 150$ GeV.
They restrict their measurements to $|\eta_\pm| \leq 1$, where $\eta_\pm$
are the pseudorapidities of $e^\pm$ respectively.
Furthermore, the numbers quoted are
averaged over the rapidity of the lepton pair. For ease of comparison, we
shall consider the
same distribution (including the actual binning for $M$
as used in the experiment). In Fig. \ref{cdffig}, we compare the
experimental data with the theoretical curves for the SM, and those
for scalars of Type $IVa$ (see Table \ref{table})\footnote{It should be noted
that the sharpness of  the $Z$--resonance is somewhat reduced by
 the rather
coarse binning adopted in the experiment.}.
We have set here $h_L = 1$, \ie the
Yukawa coupling is of the
electroweak strength. Though the effect of the
scalar starts becoming visible at around $M \simeq 65$ GeV, it
is immediately swamped by the $Z$--pole. At large $M$ values though, the
deviation is significant
and information about scalars may be extracted.
However, as even a cursory glance at the figure would reveal,
the data at present are rather poor and any quantitative statement
would be premature. We rather wish to point out that a refinement of
measurement at the high invariant mass end ($M \gsim 150 $ GeV)
of the spectrum (along with
an increase of luminosity at the possible Tevatron upgrade) would
enhance considerably the ability to detect such scalar particles.

\begin{figure}[htb]
\vskip 8in\relax\noindent\hskip -1in\relax{\includegraphics{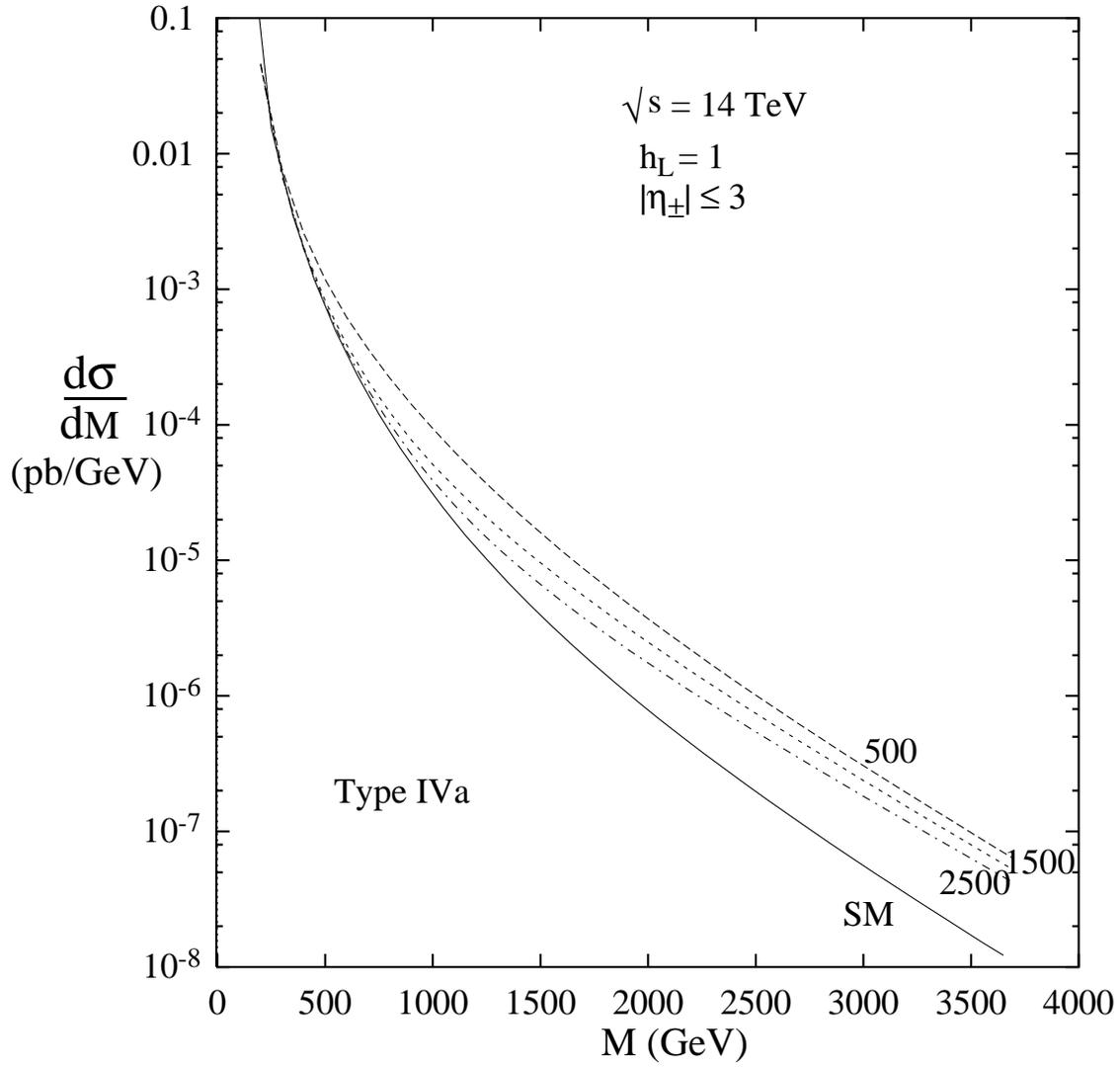}}

\vspace{-20ex}
\caption{The dilepton invariant mass distribution at the LHC. The curves
correspond to the theoretical expectation for the SM and for Type $IVa $
scalars of masses $m_\phi = 500,\:1500,\:2500$ GeV respectively.
}
       \label{massdistrib}
\end{figure}

We now turn to the LHC. Due to the large operating energy
(14 TeV) one would expect to see a more pronounced effect.
In our analysis,
we assume an integrated luminosity of 100 fb$^{-1}$. In
Fig. \ref{massdistrib}, we first compare the
invariant mass distribution for different  masses of a Type $IVa$
scalar ($h_L = 1$) with
that for the SM. We use kinematic cuts of $|\eta_\pm| \leq 3$, numbers
that have often been quoted.
As is expected, the effect of the scalar exchange is more pronounced for
large $M$. Due to the much larger energy available (and partly due
to the wider angular  coverage), the
presence of much heavier scalars can be detected.

\begin{figure}[htb]
\vskip 8in\relax\noindent\hskip -1in\relax{\includegraphics{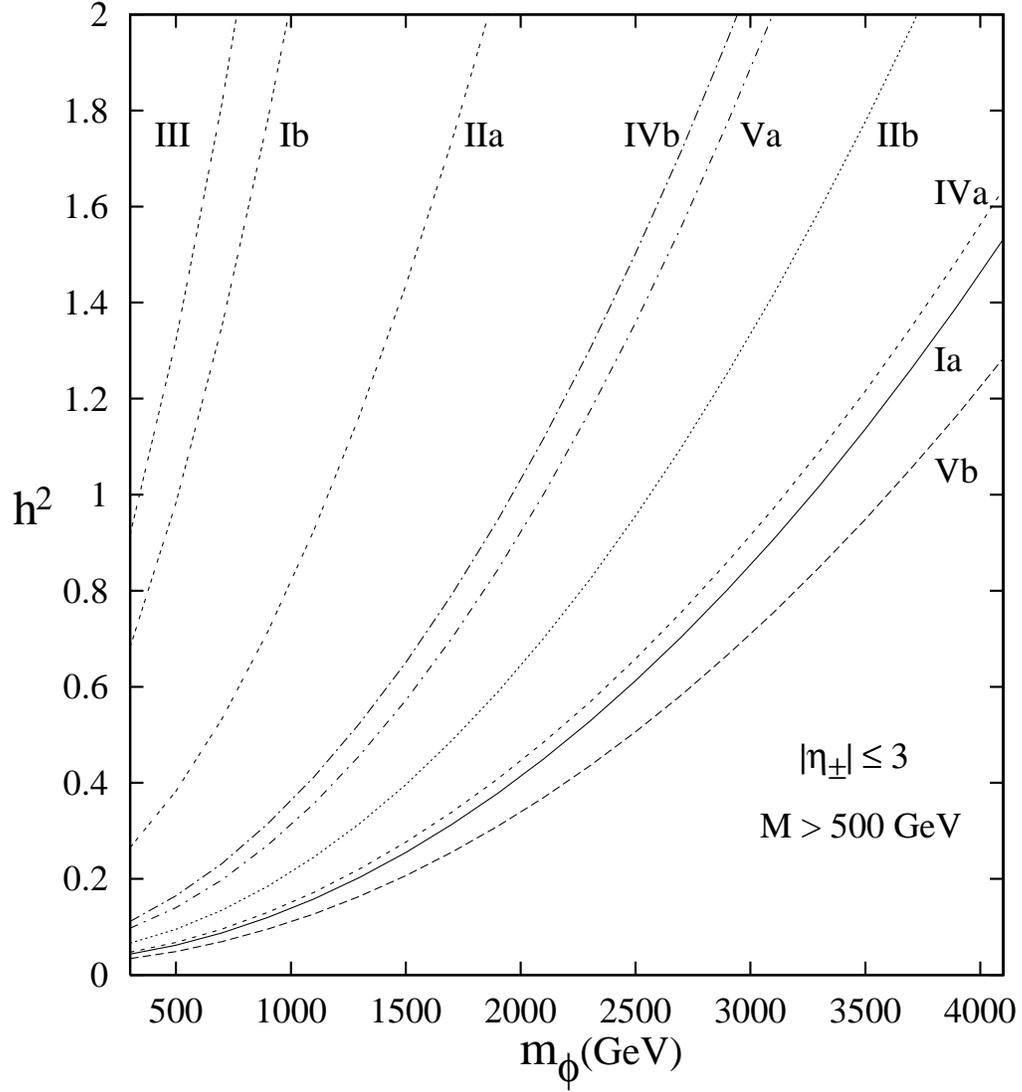}}

\vspace{-20ex}
\caption{The parameter space that can be probed at the LHC.
The different curves are for the various scalar types listed in
Table~\protect\ref{table}, with the ordinate corresponding to the respective
coupling. The part of the parameter space above the individual curves
can be ruled out at the $3 \sigma$ level.
}
       \label{mvsk}
\end{figure}

The information in
Fig. \ref{massdistrib} can be used to put bounds on the
two-parameter space  ($m_\phi, \ h$) --- here $h \equiv h_{L,R}$, as
the case may be --- that can be achieved at the LHC.
It is obvious that imposing an acceptability
cut on the invariant mass would
enhance the effect of the scalar. In our analysis, we arbitrarily set
$M_{\rm min} = 500$ GeV. Denoting the number of events expected
in the SM, and in a theory
with such a scalar by $n_{SM}$ and $n_{\phi}$ respectively, we demand that
\be
  \left| n_\phi - n_{SM} \right| \geq 3 \sqrt{n_{SM}}
        \label{3sigma}
\ee
for the effect to be considered visible. This allows us to draw contours
in the $m_\phi$--$h$ plane (Fig. \ref{mvsk}) for various types of
scalars. The region of the parameter space above the respective curves
can then be ruled out at the $3 \sigma$ level. The difference in the
sensitivity to the scalar type is a consequence of the difference in
their coupling, as also of the relative abundance of
various quarks in the proton.
While it is true that a 2.5 TeV scalar may easily be pair produced at the
LHC, such an event would afford us no handle on its couplings to the
SM fermions. Indeed,
the bounds presented here are the strongest that can be achieved on
leptoquark couplings in the near future.
For the \rp\ case, this is illustrated even better.
As we have pointed out right at
the beginning, the low energy constraints on
\rp\ couplings are stronger than those for the
usual leptoquarks.
For our interaction, \rp\ is phenomenologically identical
to a case where both Type $II\ a$ and Type $III$
scalars are present, with the
contributions adding incoherently. If we assume the two
squarks ($\tilde{u}_L$ and $\tilde{d}_R$) to be
mass--degenerate, the bound on the \rp coupling  lies in between
the two above curves, and is somewhat weaker than those derived
from low energy processes \cite{barger}. One should realise though that
this experiment provides an independent constraint on the parameter space
for the theory. Interestingly, in the extreme case of
$m(\tilde{d}_R) \gg m(\tilde{u}_L)$, the low energy constraint becomes
relatively unimportant and the experiment discussed here would provide the
strongest constraints. Indeed, for all such scalars except that of Type $III$,
dilepton data at the LHC would lead to bounds significantly stronger than
those obtained from low energy processes such as $\pi$--decay,
charge--current--universality {\em etc}.

\begin{figure}[htb]
\vskip 8in\relax\noindent\hskip -1in\relax{\includegraphics{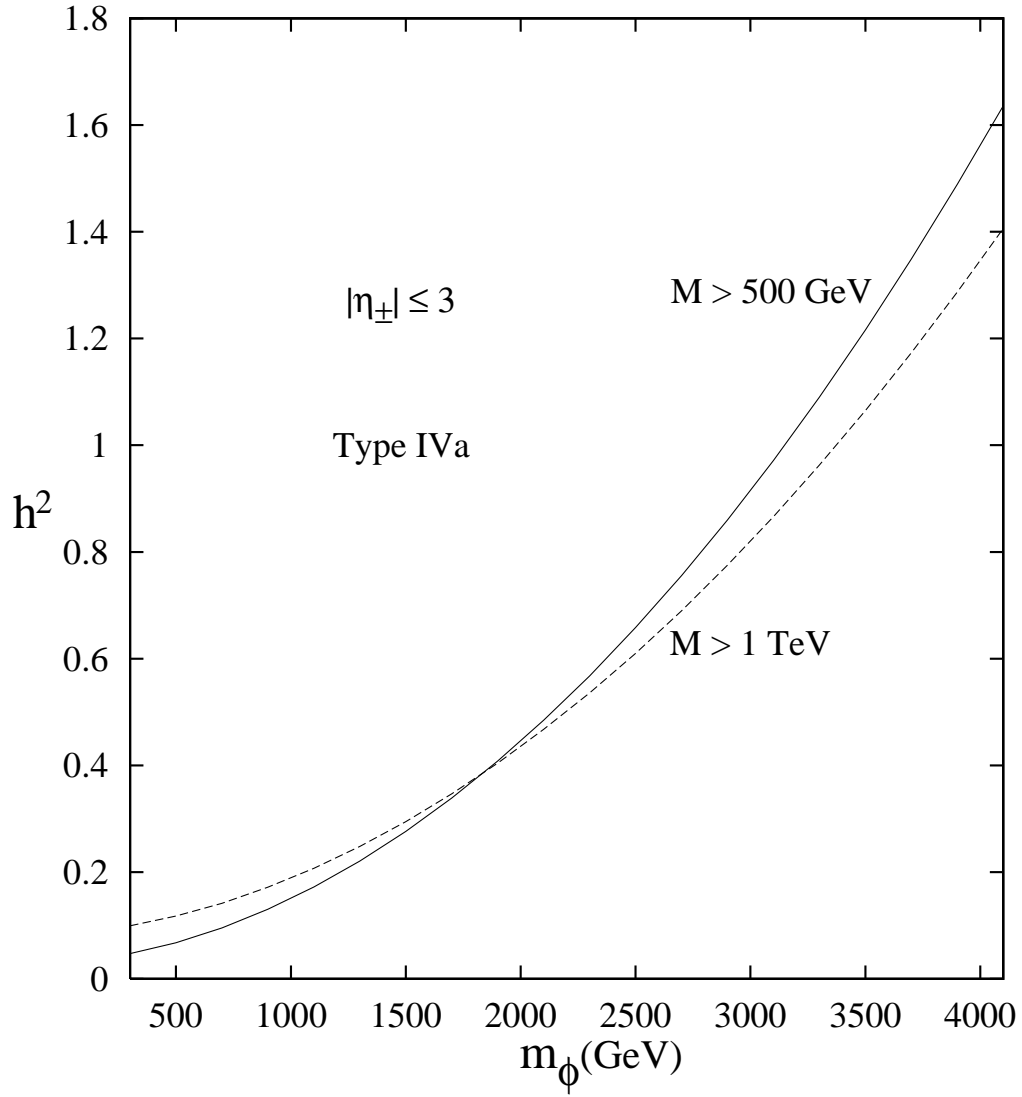}}

\vspace{-20ex}
\caption{The dependence of the exploring ability on the
invariant mass cut. The two curves ($3 \sigma$)
shown are for the Type $IVa$ scalar (see
Table~\protect\ref{table}) and for $M_{min} = 500$ GeV and 1 TeV
respectively.}
       \label{m_min}
\end{figure}

As we had mentioned, our choice of $M_{\rm min}$ was rather arbitrary.
 The exact value  that should
be adopted to maximize the effect is dependent  on the value of
$m_\phi$ that one aims to probe. On the face of it, a larger value of
$M_{\rm min}$ should serve to eliminate more of the SM contribution. This
advantage can, however, be nullified by the consequent loss of statistics.
In fact, a larger $M_{\rm min}$
is more useful for exploring larger $m_\phi$.
To illustrate our point, we superimpose the $3 \sigma$ contours for two
different values of $M_{\rm min}$ in Fig.~\ref{m_min}.

\begin{figure}[htb]
\vskip 8in\relax\noindent\hskip -1in\relax{\includegraphics{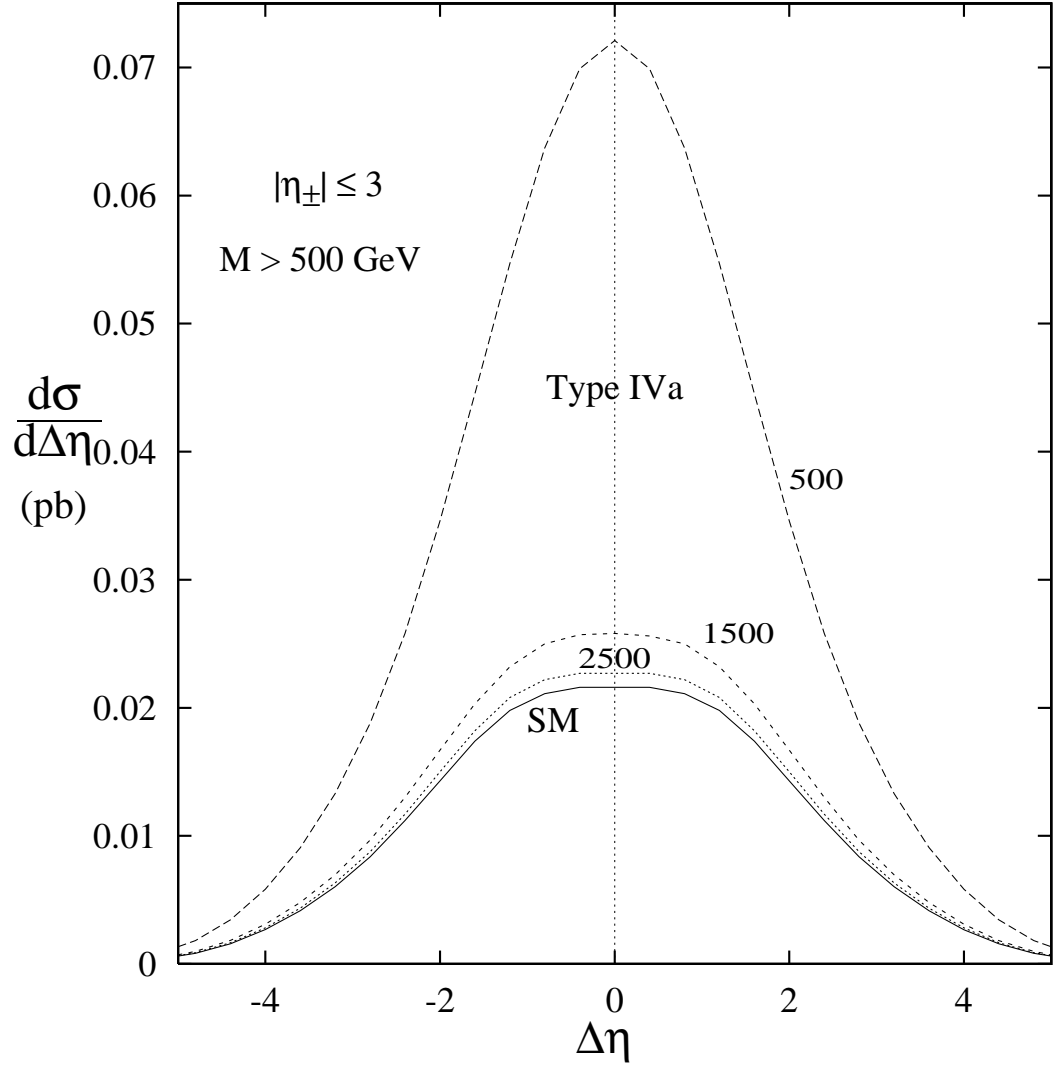}}

\vspace{-20ex}
\caption{The differential cross section for various values of
scalar masses (Type $IVa$) as a function of the difference
of the lepton rapidities.}
       \label{rapdistrib}
\end{figure}

Until now, we have neglected  the
other interesting dynamical variable in the process,
namely the difference ($ \Delta \eta = \eta_+ - \eta_-$) of the lepton
rapidities. Since this is nothing but the scattering angle in the subprocess
c.m. frame, one expects to see the difference between an $s$--channel and a
$t$--channel process. This is borne out strikingly in Fig.~\ref{rapdistrib}.
We would like to point out that instead of comparing the integrated cross
section, much more information can be gleaned from a study of the differential
distribution $d^2 \sigma/ d M \ d \Delta\eta$. This can be done, at a
quantitative level, by performing a $\chi^2$ test. We refrain from doing so,
however, as such an analysis presumes some knowledge of the detector.
We prefer instead, to only point out that once the detector parameters are
well understood, such an analysis should be
undertaken so as to improve the sensitivity of the experiment.

In summary, we have examined the virtual effect of a $t$-channel
scalar exchange in  dielectron production at CDF and tried to foresee
its possible impact at the LHC. In contrast to the pair production
mechanism, the virtual effects are very sensitive
to its Yukawa coupling. We point out that
a few  precise measurements
of the dilepton  distribution at CDF at higher invariant masses
 ($M \sim 150$ GeV) would allow us to probe the existence
 of a scalar of mass upto a
few hundred GeV and coupling the electron and a first generation quark
with electroweak strength. Similarly, at the  LHC, we observe
the possibility of probing scalar masses upto (1--3) TeV
for a Yukawa coupling of electroweak strength. These estimates are
significantly better than the bounds inferred from analyses of low--energy
processes.
We would also like to point out that such an analysis can as easily be done
for a dimuon pair, and the bounds would be similar. For a $\tau^+ \tau^-$
pair, the situation would be a bit more complicated as
$\tau$--identification
efficiency is likely to be lower. If this efficiency could be raised, such
an experiment would lead to a significant improvement in constraints on
such scalars coupling a  $\tau$ to a
quark.
Finally, one may also investigate possible couplings of leptons to
the heavier quarks, although these bounds would be weaker  on account of
the lower densities within the proton.

\newpage
\begin{table}[htb]
\vspace*{4ex}
$$
\barr{|| r | l | c | l ||}
\hline
 \multicolumn{1}{||l|}{\rm Scalar} &  &
         \multicolumn{1}{c|}{ SU(2)_L \otimes U(1)_Y} &  \\
\multicolumn{1}{||l|}{\rm Type} & \multicolumn{1}{c|}{\rm Coupling}
       & \multicolumn{1}{c|}{\rm Transformation} &
         \multicolumn{1}{c||}{\rm Remarks} \\
\hline
&&&\\
I\ a) & \overline{l_L} u_R \phi & (2,\ -7/6) &   \\
   b) & \overline{{e_R}^c} u_R \phi & (1,\ 1/3) &  \\
&&&\\
II\  a) & \overline{l_L} d_R \phi & (2,\ -1/6)
                     & \mbox{\rm Corresponds\ to}\  \tilde{u}_L   \\
   b) & \overline{{e_R}^c} d_R \phi & (1,\ 4/3) &   \\
&&&\\
III  & \overline{{l_L}^c} Q_L \phi & (1,\ 1/3)
                      & \mbox{\rm Corresponds\ to} \ \tilde{d}_R   \\
&&&\\
IV\ a) & \overline{e_R} Q_L \phi & (2, -7/6) &  \\
    b) & \overline{{l_L}^c} Q_L \phi & (3,\ 1/3) & \\
&&&\\
V\ a) & \overline{e_R} Q_L \phi & (2, -7/6) &
                   m(\phi^{5/3}) \gg m(\phi^{2/3}) \\
    b) & \overline{{l_L}^c} Q_L \phi & (3,\ 1/3)
                      & m(\phi^{1/3}) \gg m(\phi^{4/3})\\
&&&\\
\hline
\earr
$$
\caption{The possible scalar couplings between a charged lepton and
a quark, grouped according to their chiral structures. For Type $IV$,
we assume all the scalars to be mass degenerate.
Type $V$ corresponds to the (unlikely !)
case where only the coupling to $d_L$ is of importance.}
   \label{table}
\end{table}
\newpage

\end{document}